\documentclass[twocolumn,amsmath,amssymb,nofootinbib,floatfix]{revtex4}

\usepackage[pagewise]{lineno}
\usepackage{graphicx}
\usepackage{dcolumn}
\usepackage{bm}
\usepackage{hyperref}
\usepackage{color}
\usepackage{subfigure}

\newcommand{\Rmnum}[1]{\expandafter\@slowromancap\romannumeral #1@}

\def\be{\begin{equation}}
	\def\ee{\end{equation}}
\def\bea{\begin{eqnarray}}
	\def\eea{\end{eqnarray}}

\begin{document}
\title{Dual axion-like field inflation}

\author{Runchao Huang$^*$}
\author{Ruifeng Zheng$^*$}
\author{Qiaoli Yang$^\dag$}

\affiliation{Physics Department, College of Physics and Optoelectronic Engineering, Jinan University, Guangzhou 510632, China}

\begin{abstract}
Cosmic inflation is one of the most important paradigms in modern cosmology. In its simplest form, inflation is driven by a single inflaton field. However, multi-field inflation has become increasingly attractive because it can solve many theoretical and observational challenges. In this paper, we propose a particular model involving two axion-like fields with simply monodromy-dominated potentials. We demonstrate that this model is consistent with current cosmological observations.
\end{abstract}

\maketitle	
\def\thefootnote{*}\footnotetext{co-first author}\def\thefootnote{\arabic{footnote}}
\def\thefootnote{\dag}\footnotetext{corresponding author}\def\thefootnote{\arabic{footnote}}
\section{INTRODUCTION}\label{(1)}
\par
Effective field theories are one of the foundations of building our understanding of the universe. An important example is the inflation theory \cite{Guth:1980zm, Linde:1981mu, Starobinsky:1980te, Cheung:2007st}, which was proposed to solve the problems of the original Big Bang model. After the graceful exit problem was solved and later developments \cite{Linde:1983gd, Maartens:1999hf, Futamase:1987ua, Kaloper:2008fb}, inflation became a paradigm in cosmology. Many more extensions have emerged since then, giving inflation theory significant advances \cite{Armendariz-Picon:1999hyi, Linde:1993cn, Freese:1990rb, Baumann:2009ds, Dvali:1998pa}. On the observational front, evidence from the cosmic microwave background to large-scale structures has provided strong support for the paradigm.
However, the exact nature of the inflaton remains unsolved. Recently, multi-field inflation models have become increasingly attractive because these models can explain issues such as non-Gaussianities and primordial black holes \cite{Silk:1986vc, Kawasaki:1997ju, Inomata:2018cht, Bernardeau:2002jy}, where single-field inflation may struggle. Moreover, multi-field inflation could have fundamental ties to string theory \cite{Linde:1991km, Kachru:2003sx, Silverstein:2008sg}, where many axion-like and moduli fields naturally exist, so some of them could be natural candidates to drive inflation. Ultimately, cosmological observations will determine the final answer.

Beyond effective field theory, top-down theoretical approaches also offer profound insight into inflation theory. A large number of scalar and pseudo-scalar fields corresponding to different compactification modes of extra dimensions exist in string theory. The masses and couplings of these fields could span many orders in the parameter space. It would be plausible that some of them contributed to inflation in the early universe. Indeed, inflation of multiple axion-like fields has already been widely discussed.

In this paper, we propose a particular, simple scenario in which two axion-like fields with a straightforward, particle-like potential drive inflation. The potential could be natural from the perspective of particle physics. Currently, axion-like particles (ALPs) in inflation, such as N-flation \cite{Dimopoulos:2005ac}, feature periodic potentials that originated from instanton effects and possess a shift symmetry. The shift symmetry typically protects the flatness of the inflaton. Axion monodromy is another interesting possibility \cite{Silverstein:2008sg} where shift symmetry breaking leads to a monodromy effect, which allows the axion field to traverse large distances in field space, essential for driving inflation. Here we suggest two axion fields with monodromy-like potentials, assuming the periodic part is negligible compared to the quadratic term. This is different from the typical scenarios. In addition, the coupling is simply biquadratic, which presented as the primary driver of the end of inflation within the model itself. The total potential form is simple and thus could offer a tractable minimal model. Finally, we demonstrate that this model is consistent with current observational constraints.

\section{theoretical perspectives and numerical results}\label{(1)}
\par
Currently, string theory-motivated single-field inflation, such as the D-brane model, is consistent with observations. However, multiple-field models are drawing increasing attention. In this paper, we propose a scenario that two ALP fields drive inflation. The Lagrangian can be expressed as:
\begin{equation}
\begin{aligned} \label{Laa}
\mathcal{L}_{\phi\chi}= & \frac{1}{2}\partial_{\mu}\phi\partial^{\mu}\phi - \frac{1}{2}m_{\phi}^{2}\phi^{2} + \frac{1}{2}\partial_{\mu}\chi\partial^{\mu}\chi - \frac{1}{2}m_{\chi}^{2}\chi^{2}\\
&+ \frac{1}{4\varepsilon_{\phi}\varepsilon_{\chi}f_{\phi}f_{\chi}}\phi^{2}\chi^{2}~,
\end{aligned}
\end{equation}
where $m_{\varphi }$ and $m_{\chi}$ are the masses of the $\varphi$ field and the $\chi$ field; $\varepsilon_{\varphi} $ and $\varepsilon_{\chi} $ are their respective coupling coefficients. $f_{\varphi} $ and $f_{\chi} $ are the decay constants of the $\varphi$ and $\chi$ fields. The last term represents the interaction between the two fields. The field potential is then:
\begin{equation}
\begin{aligned} \label{potential}
V(\phi, \chi) &= -\frac{1}{2}m_{\phi}^{2}\phi^{2}-\frac{1}{2}m_{\chi}^{2}\chi^{2} +\frac{1}{4\varepsilon_{\phi}\varepsilon_{\chi}f_{\phi}f_{\chi}}\phi^{2}\chi^{2}.
\end{aligned}
\end{equation}
\begin{figure}[!htbp]
	\centering
	\includegraphics[scale=0.30]{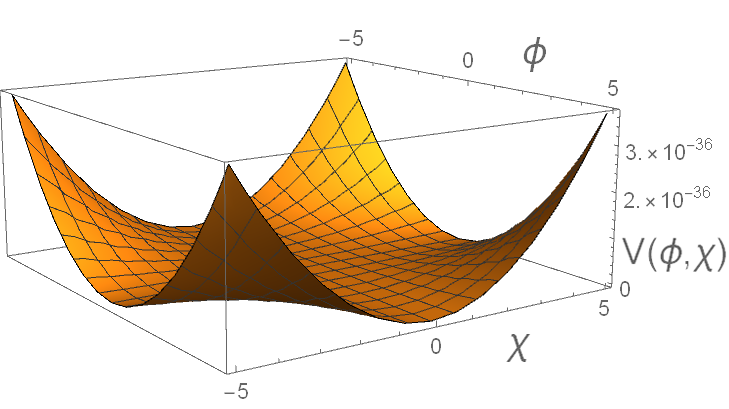}
	\caption{The field potential configuration of the proposed scenario. The yellow surface corresponding to Eq.\eqref{potential} and a natural surge due to the field coupling can be seen.}
	\label{Fig6}
\end{figure}
Figure (\ref{Fig6}) illustrates the potential energy landscape. As the field values increase, the potential maintains a relatively flat profile, but the coupling rises sharply. This indicates the coupling term being the dominant driver of inflationary dynamics.\par

Here, the masses are chosen as $m_{\phi} \sim 10^{-8}\mathrm{eV}$ and $m_{\chi} \sim 10^{-7}$$\mathrm{eV}$ for a scenario demonstration. To ensure symmetry between the fields $\phi$ and $\chi$, we implement: $\varepsilon_{\phi} \sim \varepsilon_{\chi} \sim 50$ $\mathrm{GeV}^{-1}$ and $f_{\phi} \sim f_{\chi} \sim 10^{17}$ $\mathrm{GeV}$. During inflation, it was assumed that only the two axion-like fields dominated, while the other fields had negligible effects.

\par
In the flat Friedmann-Robertson-Walker (FRW) universe, \( g_{\mu \nu} = \operatorname{diag}\left\{-1, a^{2}(t), a^{2}(t), a^{2}(t)\right\} \), where \( a(t) \) is the scale factor. Let us define \( M_{pl} = 1 \), then the dual axion-like field energy density \( \rho \) is
\begin{equation}
\begin{aligned} \label{rho}
\rho &= \frac{1}{2}\dot{\phi}^2 + \frac{1}{2}\dot{\chi}^2 + V(\phi,\chi)~,
\end{aligned}
\end{equation}
and the pressure \( p \) is \begin{equation}
\begin{aligned} \label{p}
p &= \frac{1}{2}\dot{\phi}^2 + \frac{1}{2}\dot{\chi}^2 - V(\phi,\chi)~.
\end{aligned}
\end{equation}
Denote the Hubble parameter \( H = \dot{a}(t)/a(t) \). The Friedmann equation, which determines the space-time, is
\begin{equation}
\begin{aligned}\label{friedmann}
3 H^{2} &= \frac{1}{2} \dot{\phi}^{2} + \frac{1}{2} \dot{\chi}^{2} + V(\phi, \chi)~.
\end{aligned}
\end{equation}
The equation of motion for the \( \phi \) field is
\begin{equation}
\begin{aligned}\label{EMPHI}
\ddot{\phi} + 3H\dot{\phi} + V_{\phi}(\phi,\chi) &= 0~,
\end{aligned}
\end{equation}
and for the \( \chi \) field is
\begin{equation}
\begin{aligned}\label{EMCHI}
\ddot{\chi} + 3H\dot{\chi} + V_{\chi}(\phi,\chi) &= 0~,
\end{aligned}
\end{equation}
where \( V_{\phi}(\phi,\chi) = {\mathrm{d}V(\phi,\chi)}/{\mathrm{d}\phi} \), \( V_{\chi}(\phi,\chi) = {\mathrm{d}V(\phi,\chi)}/{\mathrm{d}\chi} \). For the scalar fields to induce a sustained inflation, the potential energy needs to be much larger than their kinetic energy. In addition, one has 1) the field velocity norm needs to be much smaller than one; 2) the covariant Hessian of the potentials needs to be much smaller than one; and 3) the slow-turn condition needs to be satisfied thus the field trajectory can avoid abrupt changes.

Because the field potential dominate over the kinetic energy density: $\dot{\phi }\ll2V(\phi,\chi) ,~\dot{\chi } \ll 2V(\phi,\chi)$, equation (5) simplifies to
\begin{equation}
\begin{aligned}
H^{2}\simeq\frac{1}{3}V(\phi, \chi)~.
\end{aligned}
\end{equation}
\begin{figure}[!htbp]
	\centering
	\includegraphics[scale=0.28 ]{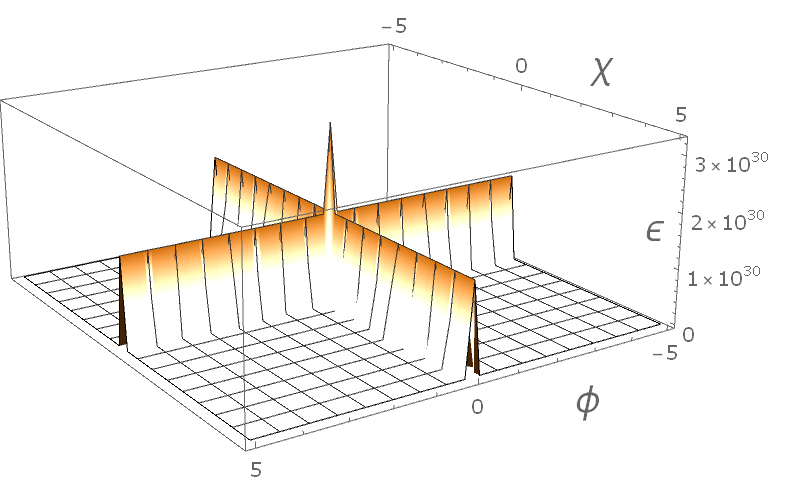}
	\caption{The evolution of the slow roll parameter $\epsilon$ in the dual-axion configuration space.}
	\label{Fig1}
\end{figure}
When the potential is flat enough, the universe experiences rapid inflation. Equation (6) and (7) can be written as: \begin{equation}
\begin{aligned}\label{EMPHI}
3H\dot{\phi}\simeq -V_{\phi}(\phi,\chi)~,
\end{aligned}
\end{equation}
\begin{equation}
\begin{aligned}\label{EMCHI}
3H\dot{\chi}\simeq -V_{\chi}(\phi,\chi)~.
\end{aligned}
\end{equation}
These equations of motion ensure the prolonged drive. One can obtain the slow roll parameters $\epsilon_{\phi\chi}$ and $\eta_{\phi\chi}$ as:
\[\epsilon = \frac{1}{2} \left( \frac{V'(\phi, \chi)}{V(\phi, \chi)} \right)^2,\]
\[\eta = \frac{V''(\phi, \chi)}{V(\phi, \chi)}.\]
The $\phi$ and $\chi$ potentials of mass and two-field coupling give rise to:
\[V'_{\phi}(\phi,\chi) = -m^{2}_{\phi}\phi + \frac{\chi^{2}}{2\varepsilon_{\phi}\varepsilon_{\chi}f_{\phi}f_{\chi}}\phi~,\]
\[V'_{\chi}(\phi,\chi) = -m^{2}_{\chi}\chi + \frac{\phi^{2}}{2\varepsilon_{\phi}\varepsilon_{\chi}f_{\phi}f_{\chi}}\chi~,\]
\[V_{\phi\phi}^{\prime\prime}(\phi, \chi) = -\,m_{\phi}^{2} + \frac{\chi^{2}}{2\varepsilon_{\phi}\varepsilon_{\chi}f_{\phi}f_{\chi}}~,\]
\[V_{\phi\chi}^{\prime\prime}(\phi, \chi) = \frac{\phi\chi}{\varepsilon_{\phi}\varepsilon_{\chi}f_{\phi}f_{\chi}}~,\]
\[V_{\chi\chi}^{\prime\prime}(\phi,\chi) = -m_\chi^2 + \frac{\phi^2}{2\varepsilon_\phi\varepsilon_\chi f_\phi f_\chi}~,\]
\[V_{\chi\phi}^{\prime\prime}(\phi,\chi) = \frac{\phi\chi}{\varepsilon_\phi\varepsilon_\chi f_\phi f_\chi}~.\]
The respective slow roll parameter of field velocity norm $\epsilon$ is
\begin{equation}
\epsilon= \frac{1}{2}\frac{\left(
m_{\phi}^{2}\phi - \frac{\chi^{2}\phi}{2\varepsilon_{\phi}\varepsilon_{\chi}f_{\phi}f_{\chi}} \right)^2 +\left( m_{\chi}^{2}\chi
- \frac{\phi^{2}\chi}{2\varepsilon_{\phi}\varepsilon_{\chi}f_{\phi}f_{\chi}}\right)^2}{
\left( \frac{1}{2}m_{\phi}^{2}\phi^{2} + \frac{1}{2}m_{\chi}^{2}\chi^{2}
- \frac{1}{4\varepsilon_{\phi}\varepsilon_{\chi}f_{\phi}f_{\chi}}\phi^{2}\chi^{2}\right)^2~.
}
\end{equation}

\begin{figure}[!htbp]
	\centering	\includegraphics[scale=0.30]{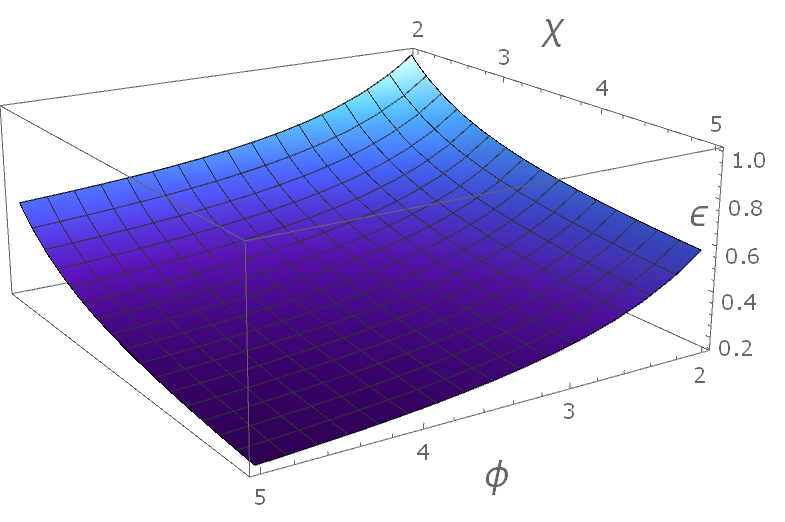}
	\caption{The evolution of the slow roll parameter \( \epsilon \) in the first quadrant of $\phi$ and $\chi$. The $\phi$ and $\chi$ will decrease until $\epsilon=1$.}
	\label{Fig2}
\end{figure}

Figure (\ref{Fig2}) shows the evolution behavior of the slow roll parameter \( \epsilon \) with \( \phi \) and \( \chi \). The \( \epsilon \) value remains extremely low in the early stage of inflation (satisfying the slow roll condition), and gradually increases with the weakening of potential energy.
\par
\begin{figure}[!htbp]
	\centering
	\includegraphics[scale=0.5]{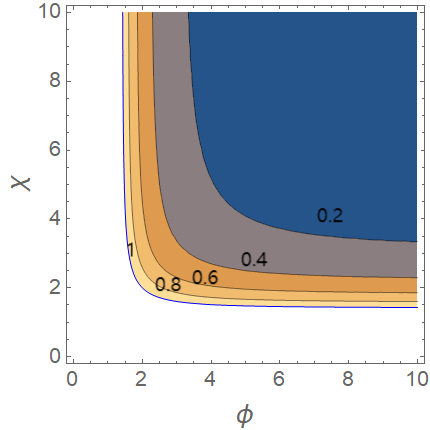}
	\caption{The parameter space of \( \phi \) and \( \chi \) associated with the evolution of the slow roll parameter \( \epsilon \).}
	\label{2D}
\end{figure}

When inflation ends, \( \epsilon = 1 \) with the corresponding \( \phi_e \) and \( \chi_e \). Figure (\ref{2D}) shows the respective \( \phi \) and \( \chi \), where the outermost line corresponds to \( \epsilon = 1 \).
\par
The second slow roll parameter \( \eta \) can be expressed as
\begin{equation}
    \eta=\frac{m_\phi^2+m_\chi^2-\frac{\phi^2+\chi^2}{2\varepsilon_\phi\varepsilon_\chi f_\phi f_\chi}-\frac{2\chi\phi}{\varepsilon_\phi\varepsilon_\chi f_\phi f_\chi}}{\frac{1}{2}m^2_\phi\phi^2+\frac{1}{2}m_\chi^2\chi^2-\frac{1}{4\varepsilon_\phi\varepsilon_\chi f_\phi f_\chi}\phi^2\chi^2}.
\end{equation}
Similar to Figure (\ref{Fig2}), Figure (\ref{Fig4}) shows the evolution behavior of the slow roll parameter \( \eta \) with respect to \( \phi \) and \( \chi \). The contribution of the coupling term to \( \eta \) is significant in the later stage of inflation. Figure (\ref{eta-2D}) shows the critical line (solid blue line) of \( \eta = 1 \), which determines the end of inflation.
\begin{figure}[!htbp]
	\centering
	\includegraphics[scale=0.25]{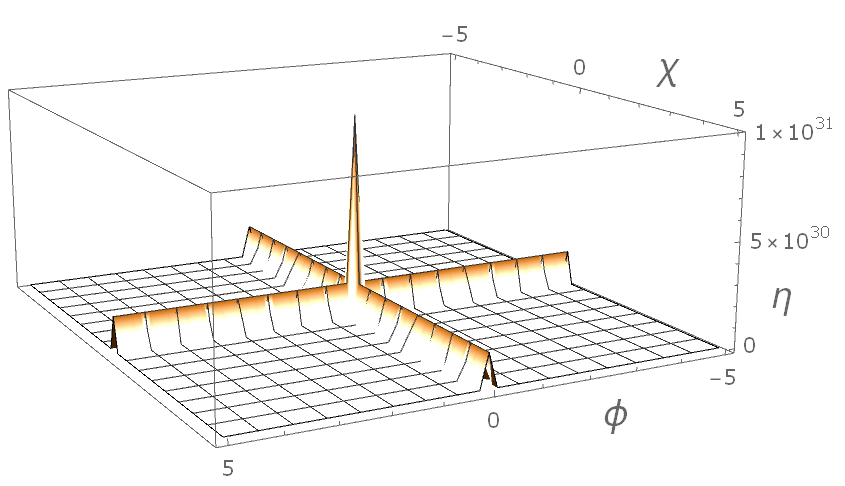}
	\caption{The evolution of the slow roll parameter \( \eta \) with \( \phi \) and \( \chi \) configuration space.}
	\label{Fig3}
\end{figure}
\begin{figure}[!htbp]
	\centering
	\includegraphics[scale=0.25]{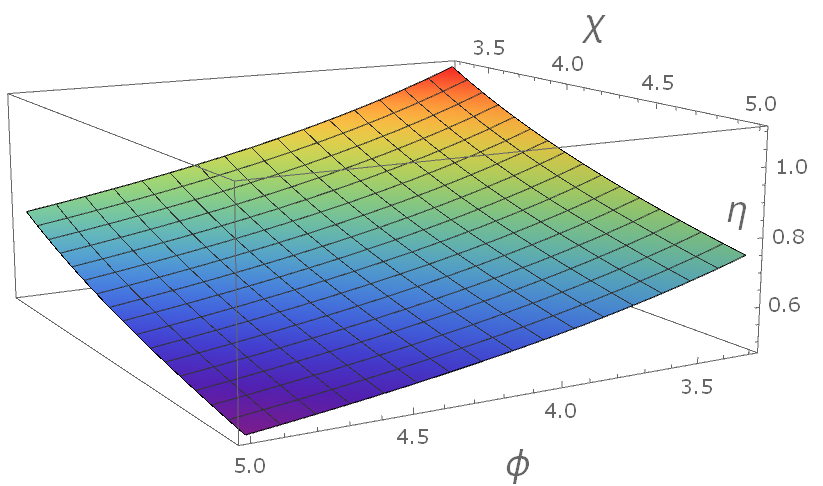}
	\caption{The evolution of the slow roll parameter \( \eta \), with \( \phi \) and \( \chi \) gradually evolving from larger values to smaller values. \( \eta = 1 \) indicates the end of inflation.}
	\label{Fig4}
\end{figure}
\begin{figure}[!htbp]
	\centering
	\includegraphics[scale=0.4]{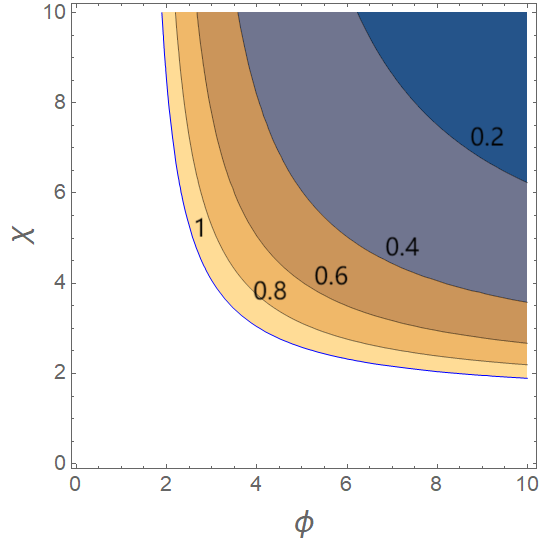}
	\caption{The parameter space of \( \phi \) and \( \chi \), corresponding to the evolution of the slow roll parameter \( \eta \); also see Figure (\ref{Fig4}).}
	\label{eta-2D}
\end{figure}

As the scale factors determine the expansion of the universe during inflation, one can introduce the \( N \) of e-foldings:
\begin{equation}
\begin{aligned}
 N(\phi_e, \chi_e, \phi_i, \chi_i)
 =& -\frac{m_{\phi}^2\varepsilon_{\phi}\varepsilon_{\chi}f_{\phi}f_{\chi}}{4} (\phi_e^2 - \phi_i^2)(\ln\chi_e - \ln\chi_i)\\
  & - \frac{m_{\chi}^2\varepsilon_{\phi}\varepsilon_{\chi}f_{\phi}f_{\chi}}{4} (\chi_e^2 - \chi_i^2)(\ln\phi_e - \ln\phi_i)\\
 &+ \frac{1}{16}(\phi_e^2 - \phi_i^2)(\chi_e^2 - \chi_i^2).
 \end{aligned}
 \end{equation}
To solve the horizon problem, the value of the e-folding number is usually between \( 50-60 \), which can be satisfied here \cite{Liddle:2000cg}. When the initial parameters of $\phi_i$ and $\chi_i$ are set to $5\mathrm{M_{pl}}-9\mathrm{M_{pl}}$, we will obtain a spectral index of approximately
$n_s\approx0.967658\pm0.012841$,
which satisfies the current cosmological observations.

\section{DISCUSSIONS AND CONCLUSIONS}\label{S5}
In the context of string theory, multiple axion and moduli fields naturally arise. Therefore, a multi-field inflation scenario could be preferred from a top-down perspective. Additionally, multi-field inflation models could address some phenomenological issues such as fine-tuning and non-Gaussianities more naturally. In this paper, we propose a two-axion-like field inflation model with straightforward monodromy-dominated potentials. Compared with most multiaxion models, the proposed model is very simple and the steep rise in the coupling term is the primary driver of the end of inflation.

By analyzing the slow-roll parameters and the spectral index, we find that the model is viable and offers a substantial parameter space for exploration. Future studies of its cosmological effects, such as primordial fluctuations and non-Gaussianity, could be worthwhile.

\section*{Acknowledgments}
We would like to thank Danning Li, Chao Niu, and Mengchao Zhang for valuable discussions. This work has been supported in part by the NSFC under Grant No.12150010.

%

\end{document}